\documentclass[journal]{IEEEtran}
\IEEEoverridecommandlockouts
\usepackage{cite}
\usepackage{amsmath,amssymb,amsfonts}
\usepackage{algorithm}
\usepackage{algorithmicx}
\usepackage{graphicx}
\usepackage{textcomp}
\usepackage{xcolor}
\usepackage[justification=centering]{caption}
\usepackage{algpseudocode	}
\makeatletter
\newcommand{\removelatexerror}{\let\@latex@error\@gobble}
\makeatother
\usepackage{comment}
\usepackage{booktabs}

\usepackage{soul}
\soulregister\cite7 
\soulregister\citep7 
\soulregister\citet7 
\soulregister\ref7
\soulregister\pageref7

\def\BibTeX{{\rm B\kern-.05em{\sc i\kern-.025em b}\kern-.08em
    T\kern-.1667em\lower.7ex\hbox{E}\kern-.125emX}}
\begin{document}

\title{Machine Learning aided Precise Indoor Positioning}

\author{ Anqi Yin and Zihuai Lin
\thanks{Anqi Yin and Zihuai Lin are with the School of Electrical and Information Engineering, The University of Sydney, Australia (e-mail: zihuai.lin@sydney.edu.au }
}

\maketitle

\begin{abstract}
This study describes a UWB and Machine Learning (ML)-based indoor positioning system. We propose a simple mathematical strategy to create data to reduce the job of measurements for fingerprint-based indoor localization systems. A considerable number of measurements can be avoided this way. The paper compares and contrasts the performance of four distinct models. Most test locations' average error may be reduced to less than $150$ mm using the best model.
\end{abstract}




\maketitle
\section{Introduction}
\label{sec:introduction}
Recently, Internet of Things (IoT) has captured much attention in academics and industries. With the trend of the smartphone, the localization system comes into people’s daily life and plays a more and more critical role.
There are mainly two categories of localization: Outdoor localization and indoor localization. For the outdoor positioning system such as the GPS, it can return a satisfactory location result in more than 100 countries if the user is in an outdoor environment. However, in terms of indoor positioning, the GPS faces some huge disadvantages such as the satellite signal cannot penetrate the wall, or the error could be even larger than the indoor space.

Indoor positioning can be used in many applications nowadays. For example, in a hospital, it could be useful to get the current location of each staff and patient. In this way, it could be easier to manage patients and mobilize manpower to deal with an emergency issue as soon as possible. The current technological development makes satellite positioning excellent in outdoor scenes. However, in terms of indoor positioning, satellite positioning with an accuracy of less than one meter is difficult to achieve at an affordable cost at the same time \cite{1}. 

Ultra-Wideband (UWB) technology can be employed in indoor positioning because it can perform high-accuracy time measurement. In comparison to an indoor GPS system, UWB offers the advantage of allowing non-line-of-sight positioning \cite{2,leng2020implementation}. Because the UWB has a high time resolution, it can be utilized for a variety of measurements. Time of Arrival (TOA), Time Difference of Arrival (TDOA), Angle of Arrival (AOA), Received Signal Strength Indication (RSSI) \cite{position2}, and other methods \cite{position1} are commonly used. Among those methods, TDOA can reach centimetre-level precision \cite{3}. Besides the above mentioned methods, emerging radio sensing techniques, such as \cite{GI1,GI2,GI3,GI4} can also be used for indoor positioning.
Indoor positioning can be done by using the fingerprint method. It will create a database with accurate point location data, requiring the use of known points as references, and collecting data sent from the target point \cite{4}. The data will then be compared to the database to determine the best match and location.

Machine learning is widely employed in a variety of fields, and it could be applied to increase indoor location accuracy. Machine learning is used in the fingerprint placement approach to classify the points and improve accuracy in order to discover the best match \cite{5}. K-Nearest Neighbor (KNN) classification method, Decision Tree algorithm, Random Forest algorithm, Support Vector Machine (SVM), and Multilayer Perceptron are some of the common algorithms used in fingerprint positioning.

 In this paper, we develop localization techniques to improve the accuracy of UWB-based indoor positioning using Machine learning. This paper uses decision trees, random forests, KNN and soft voting to determine the locations, among which the soft voting algorithm performs best. The paper also introduce four different models for machine learning. 

The remainder of the paper is organized as follows. Section \ref{sec:II} introduces the background about indoor positioning and the related techniques. Section \ref{sec:III} reviews the existing methods of indoor positioning based on machine learning. Section \ref{sec:IV}  introduces the methods used in the paper. Section \ref{sec:V} describes the process of the experiment. Section \ref{sec:VI} analyzes the experimental results. The last Section gives a summary of this work and the future work plan.

\section{Related work}\label{sec:II}

A Bluetooth-based indoor positioning approach with machine learning is introduced in \cite{20}. The authors develop a machine learning algorithm to improve the fingerprinting positioning technology. The RSSI value in each mapping point is trained using a machine learning algorithm. After the training section, it predicts the perfect match between the database and the test RSSI value. After finding the best match, the computer locates the test point based on the related training information. The accuracy of positioning is highly improved in the experiment. 

The authors of \cite{21} develop a machine learning algorithm for a UWB-based indoor positioning system to improve the accuracy. In that paper, Gaussian Mixture Model and Random Forest algorithm are used to locate static targets or moving targets. It is shown that the Random Forest algorithm has a better performance than the Gaussian Mixture Model in terms of accuracy, specificity, and sensitivity.

The authors of \cite{5} compare the performance of various machine learning algorithms. The algorithms they used for comparison are KNN, Random Forest, SVM and Multilayer Perceptron. The result is shown in Table \ref{table:1}, which shows that Random Forest has the highest accuracy for the investigated indoor positioning system. 

\begin{table}[h!]
\renewcommand{\arraystretch}{1.5}
\centering
\begin{tabular}{|c c|} 
 \hline
Algorithm  & Accuracy(m) \\ 
 \hline
 Random Forest & 2.23 \\  	
 KNN & 2.27 \\ 
 SVM & 2.43 \\
 Multilayer Perceptron & 2.49 \\
 \hline
\end{tabular}
\caption{Accuracy for ML Algorithms}
\label{table:1}
\end{table}

In \cite{5}, the authors also add a particle filter (PF) in the algorithm to achieve high accuracy. The result is shown in Table \ref{table:2}. From the result, we can see that the accuracy is improved significantly using particle filters in all four machine learning algorithms. Among them, the combination of the Random Forest and particle filter achieves the highest accuracy.

\begin{table}[h!]
\renewcommand{\arraystretch}{1.5}
\centering
\begin{tabular}{|c c|} 
 \hline
Algorithm  & Accuracy(m) \\ 
 \hline
 Random Forest + PF & 1.65\\  	
 SVM + PF & 1.68 \\ 
 Multilayer Perceptron + PF & 1.70 \\
 KNN + PF & 1.83 \\
 \hline
\end{tabular}
\caption{Accuracy for ML Algorithms with PF}
\label{table:2}
\end{table}

In \cite{22}, a Naive Bayes (NB) machine learning algorithm is implemented to enhance the accuracy. This algorithm  focuses on the UWB Indoor Positioning Services. Receiving Operating Curves (ROC)s are used to evaluate the performance of the algorithm. It is shown that as the distance between the target point and the reference point grows, the inaccuracy between the measured value and the real value grows. The experiment is carried out in both line-of-sight and non-line-of-sight conditions. The results show that the NB algorithm's classification effect can greatly increase indoor positioning accuracy, and the technique can be used in both line-of-sight and non-line-of-sight scenarios.

In \cite{23}, a non-line-of-sight ultra-bandwidth positioning technique is developed for non-line-of-sight systems. To boost accuracy, the system employs machine learning techniques. In the anechoic chamber and the underground corridor, two machine learning techniques, Fisher's linear discriminant and support vector machine (SVM), are utilized. The SVM method can obtain 92 percent accuracy in an anechoic room setting, while Fisher linear discrimination can achieve close to 100 percent accuracy. The application of non-line-of-sight signal positioning is very useful in the industrial field \cite{24}. For stationary objects, using the power characteristics of the received signal is sufficient for tracking. However, for moving objects, the accuracy of the power characteristics is insufficient. A machine learning classifier using the multi-layer perceptron (MLP) and enhanced decision tree (BDT) to improve accuracy is developed in \cite{23,24}. It is shown that BDT can increase the accuracy from $79\%$ to $87\%$.

\section{Methods} \label{sec:III}

\subsection{Devices Introduction}

In this work, the hardware used to measure UWB signals in this experiment is DW1000FOLLOWER. 
Four pieces of hardware are employed in the experiment, three of which function as anchor nodes and one as a target node. One of the anchor nodes is utilized as a transceiving node for receiving measurement results and connecting to a computer.  The DW1000FOLLOWER system supports the IEEE802.15.4-2011 protocol and TOA based distance measurement with a transmission distance of up to 300 meters. 





\subsection{The Trilateration Algorithm}

\begin{figure}[htp]
    \centering
    \includegraphics[width=10cm]{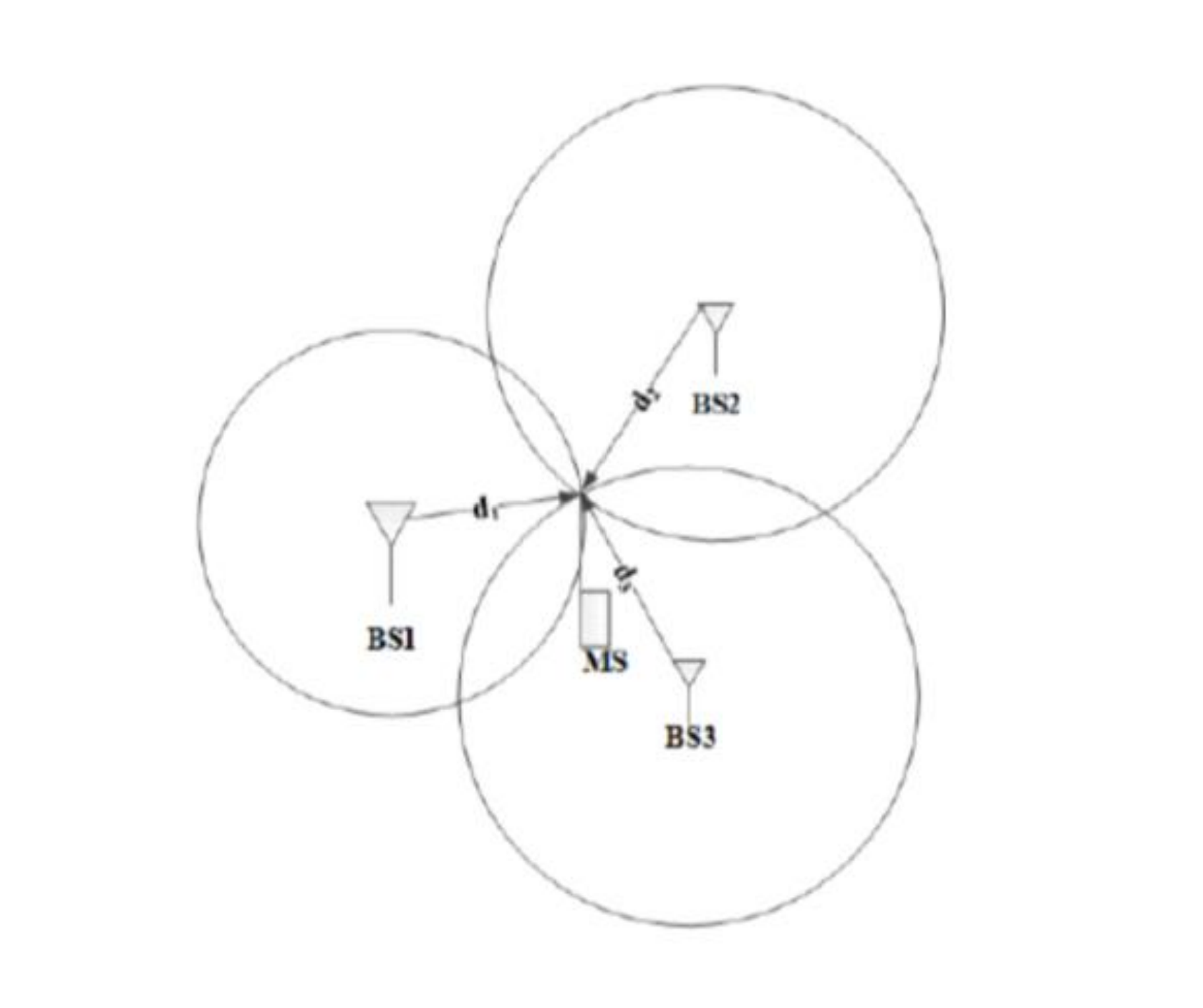}
    \caption{Trilateration algorithm method \cite{25}}
    \label{fig:3.4}
\end{figure}

The Trilateration algorithm is a commonly used method for calculating the target node location \cite{25}. The principle of the algorithm is shown in Fig. \ref{fig:3.4}. There are three non-collinear base stations BS1, BS2, BS3 and an unknown terminal MS on a plane, and the distances from the three base stations to the terminal MS have been measured as d1, d2 and d3. Then when using the base station as the center and the distance as the radius, three circles can be determined. The coordinate of the unknown node MS is the intersection points of the three circles. 

\subsection{Fingerprint Positioning}

\begin{figure}[htp]
    \centering
    \includegraphics[width=4cm]{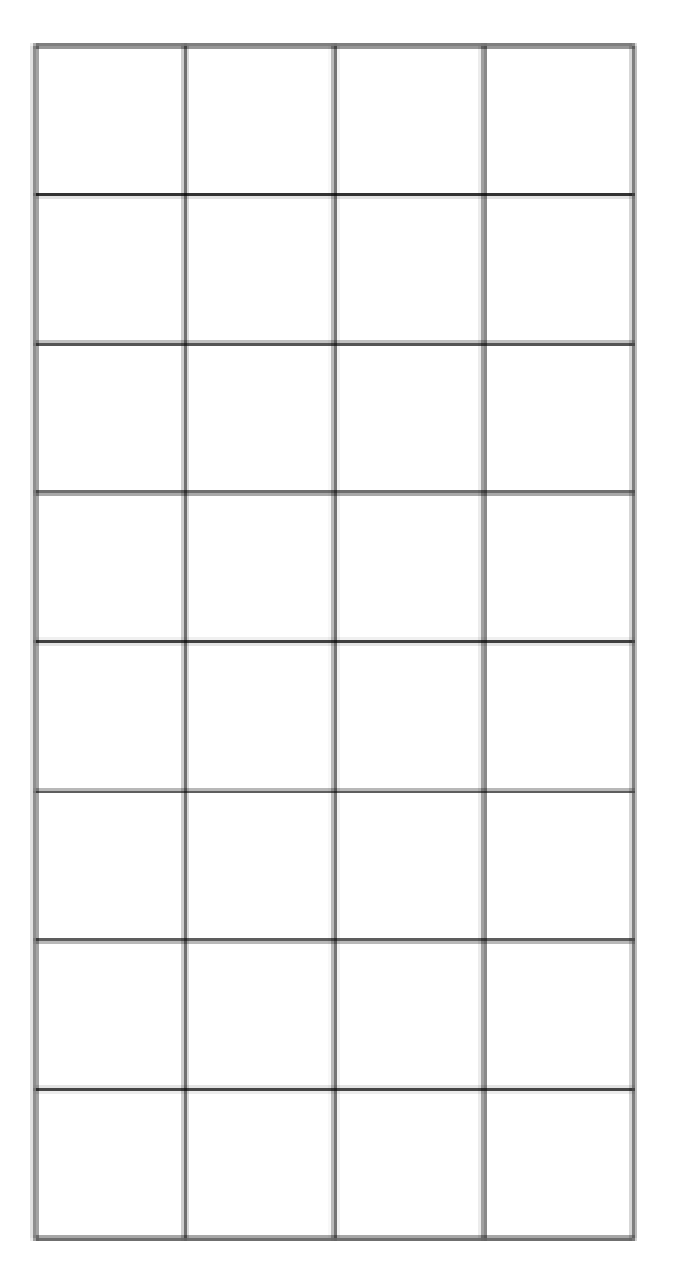}
    \caption{The way of dividing the area}
    \label{fig:3.5}
\end{figure}

In the experiment, the fingerprint positioning method is used to build a model for machine learning. In fingerprint positioning, the whole positioning area is divided into several equal parts. The model stores the location information for each part. The location information is the distances between the testing node and three anchor nodes. The machine learning algorithm classifies the location information of the testing point into each part to determine its location. For example, in Fig. \ref{fig:3.5}, the area is divided into $32$ parts. When building the model, it collects the location information for the lower-left vertex of each square. Then it finds a match location information for the test point, the coordinates corresponding to this location information are the coordinates of the test point. In the experiment, high-level accuracy is required, so it is necessary to divide the area into as smaller parts as possible. Considering the operating efficiency, in this work, we divide the entire area of two meters by one meter into $3200$ units in total. 

\subsection{Machine Learning Methods}

 Machine learning has a good performance in classification and regression. In this work, we build a machine learning model for a target localization based on the fingerprint method. After training this model and sorting out the optimal solution, the accuracy of indoor positioning can be improved. 
 

There are three main steps for our localization method based on machine learning: 

\begin{itemize}
 \item The first step is to build a model using observation (reference) points.
 \item The second step is to train the model using machine learning algorithms.
 \item The third step is to run the program with test points and get results.
\end{itemize}

In the first step, when building a model, four boundary points will be used as reference points to build a data model. These $4$ points are $(100,100)$, $(900,100)$, $(100,1900)$ and $(900,1900)$. The interval between the horizontal and vertical coordinates of each point in the model is $25$ mm. The second step is to train the model established above. The training is based on different machine learning algorithms. The third step is to calculate the data obtained from the test points to obtain the maximum error and average error.

Four types of machine learning algorithms are used in this work: Decision Tree, Random Forest, KNN and Soft Voting. The Soft Voting algorithm is based on the Decision Tree and KNN. 




\section{System model} \label{sec:IV}

\begin{figure}[htp]
    \centering
    \includegraphics[width=8cm]{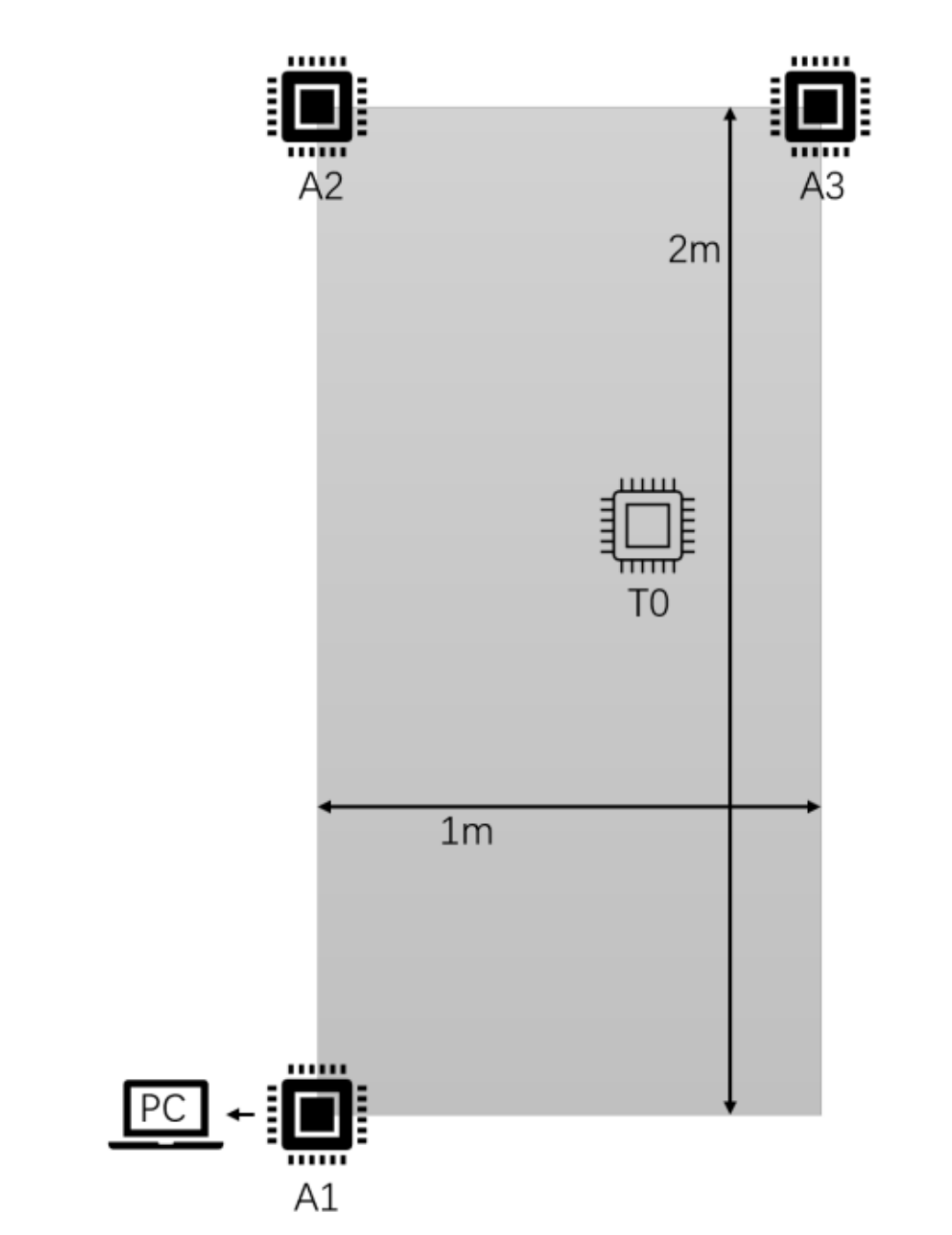}
    \caption{System module}
    \label{fig:4.1}
\end{figure}

The experiment will use three anchor nodes and one target node, one anchor node is used to transmit data at the same time. The system model is shown in Fig. \ref{fig:4.1}. This experiment is carried out in a rectangular area of one meter by two meters, and three anchor nodes, $A_1$, $A_2$, $A_3$, are placed at the three vertices of the rectangular area. The target node is represented by $T_0$.

For training, we collect location data for 10 locations, and 500 sets of data are collected for each location to provide a sufficient database for the machine learning method.

\section{Data processing} \label{sec:V}

 During the training, as the exact distance is known, it can be compared with the measured value. 

\subsection{Outlier removal}

In the data obtained from the test, one or several data are too different from other data, those data are called outliers. Outliers must be eliminated to benefit both model construction and accuracy testing. In terms of model construction, deleting irrational data improves not only the model's accuracy but also its universality. For the follow-up accuracy test, removing outliers can get rid of the negative influence of unreasonable data on the experimental results, and achieve a more realistic average error.

The data from all the observation locations forms a matrix as input values. The outliers are the value that differs from the median by more than three times the converted median absolute deviation (MAD). After eliminating outliers, a new set of data is generated. In this experiment, all the following processes use the new data set as a database.

\subsection{Data correction}

Analyze the values obtained by performing $500$ repeated measurements at each location. When the real distance is less than 1000 mm, there are certain unavoidable errors between the measured value and the real value. However, when the actual distance is greater than 1000 mm, the error of the measured distance value increases significantly.

This correction method is to correct the original data to make it closer to the true value. When the measured distance value is greater than 1000 mm, this value is scaled down to try to offset the excessive error found during the experiment. Three different ratios are used in the test, they are reduced to 90\%, 85\% and 80\% of the original data. 

\section{Data Modeling} \label{sec:VI}

\begin{figure}[h!]
    \centering
    \includegraphics[width=5cm]{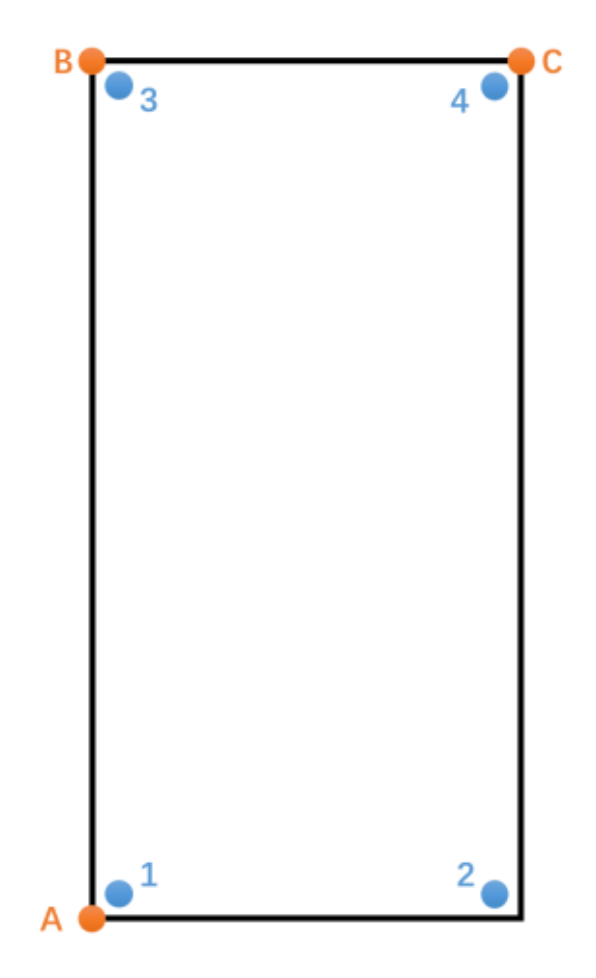}
    \caption{Anchor nodes A, B, C, and reference locations 1,2,3,4 in the area.}
    \label{fig:4.4}
\end{figure}

The collection of the location information costs time and it is impossible when having a large area with 3200 units. A method to build a model using 4 observation points is implemented. The distances from 4 observation points 1,2,3,4 to 3 anchor nodes A, B, C are measured first. Fig. \ref{fig:4.4} shows the geographic locations of four observation points and three receiving nodes.

\begin{figure}[h!]
    \centering
    \includegraphics[width=9cm]{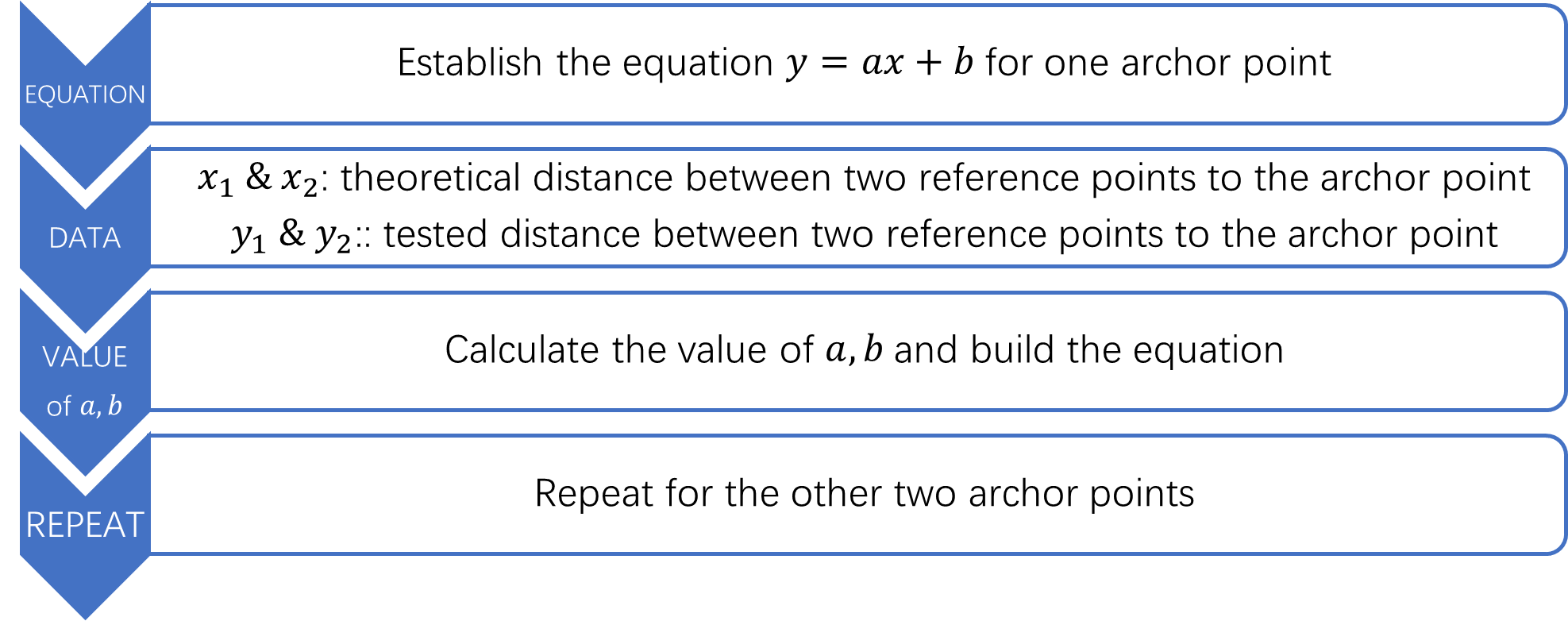}
    \caption{flowchart for building the equations.}
    \label{fig:4.5}
\end{figure}

The basis for modelling is to establish equations between the actual distance from the test location to the three anchor nodes and the distance obtained from the test. Once the equations are determined, the possible tested distance for each new location can be calculated by the equation using its theoretical distance. The establishment process of the equation is shown in Fig. \ref{fig:4.5}, in which the measurements are used to find the parameters of $a$ and $b$.

60 sets of tested data are randomly selected from the total 300 sets. Therefore, 60 sets of equations could be calculated for each anchor node. In the experiment, four types of models are established for testing.

\subsection{Model One}

\begin{figure}[h!]
    \centering
    \includegraphics[width=9cm]{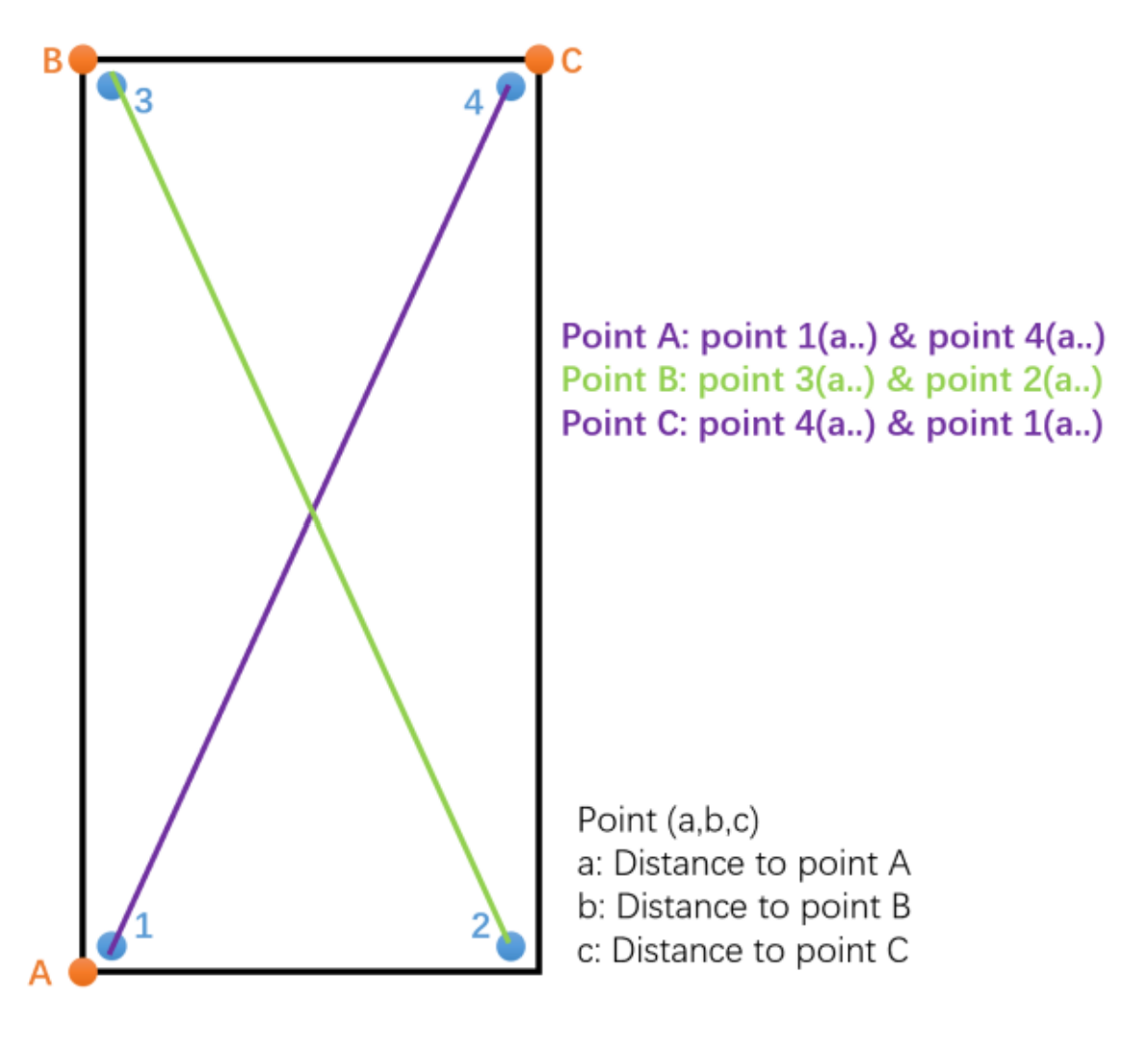}
    \caption{Algorithm diagram of Model One}
    \label{fig:4.6}
\end{figure}

For each anchor node location, the data for the two observation points on the diagonal is used for calculation as shown in Fig. \ref{fig:4.6}. 

For anchor node A, the corresponding values for the distance to node A from points 1 and 4 are used for calculation. For anchor node B, the corresponding values for the distances from node A to points 2 and 3 are used. For anchor node C, the corresponding values for the distances to node A for points 1 and 4 are used. In this way, the $a$, $b$ values for anchor nodes A and C are very similar, and the data used are derived from the data for anchor node A, which may be of limited applicability to anchor B and anchor C, so a second way to model is created.

\subsection{Model Two}

\begin{figure}[h!]
    \centering
    \includegraphics[width=9cm]{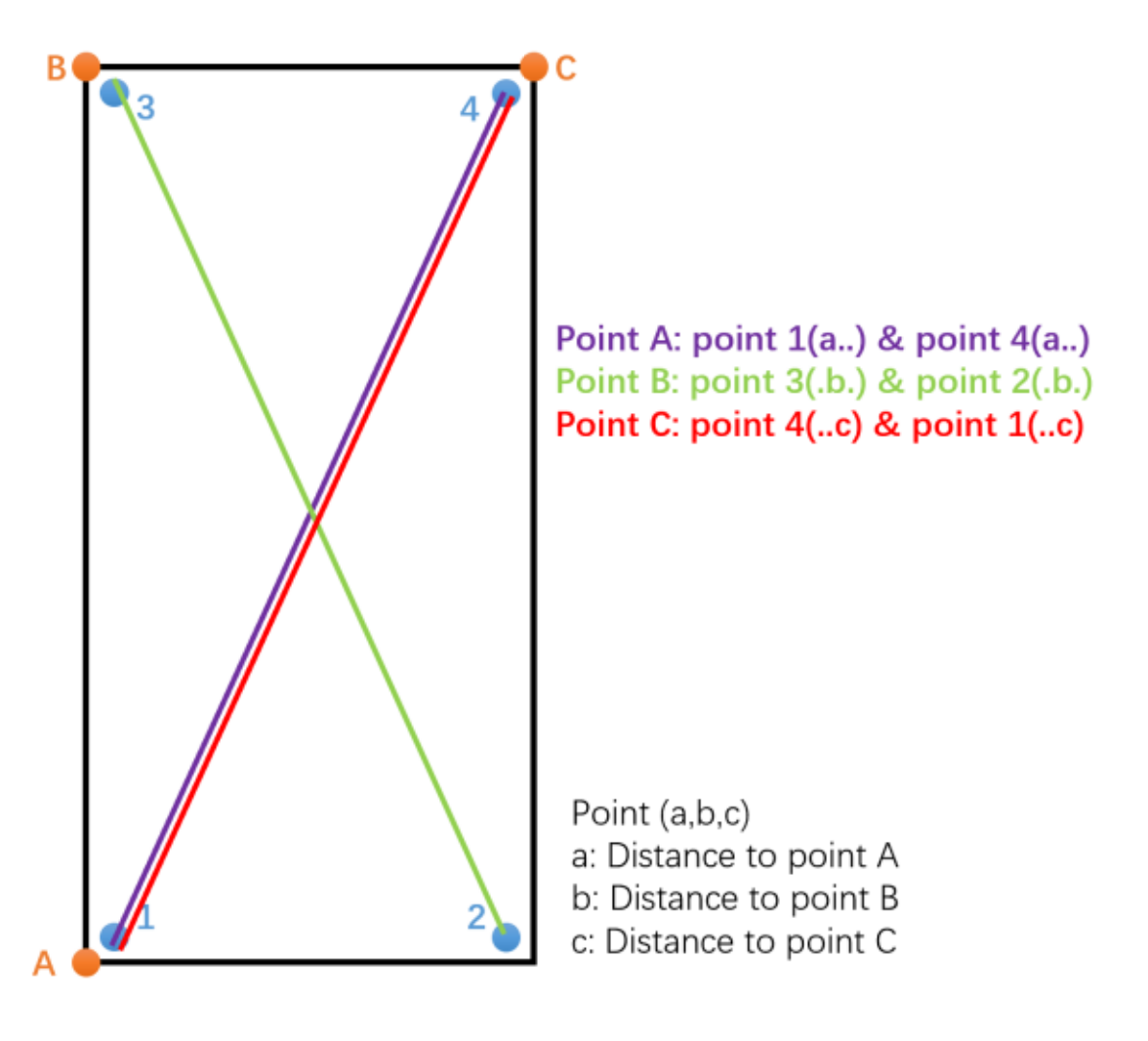}
    \caption{Algorithm diagram of Model Two}
    \label{fig:4.7}
\end{figure}

The process of Model Two is very similar to Model One. The difference between the two approaches is that in Model Two, the data used for different anchor nodes is the distance between the observation point to each anchor node itself. The rules are shown in Fig, \ref{fig:4.7}, for anchor node A, the corresponding values for the distance to node A from points $1$ and $4$ are still used. However, for anchor node B, this model will use the distances from node B to points 2 and 3. For anchor node C, the distances from node C to points 1 and 4 are used in the calculation. 

\subsection{Model Three}

\begin{figure}[h!]
    \centering
    \includegraphics[width=9cm]{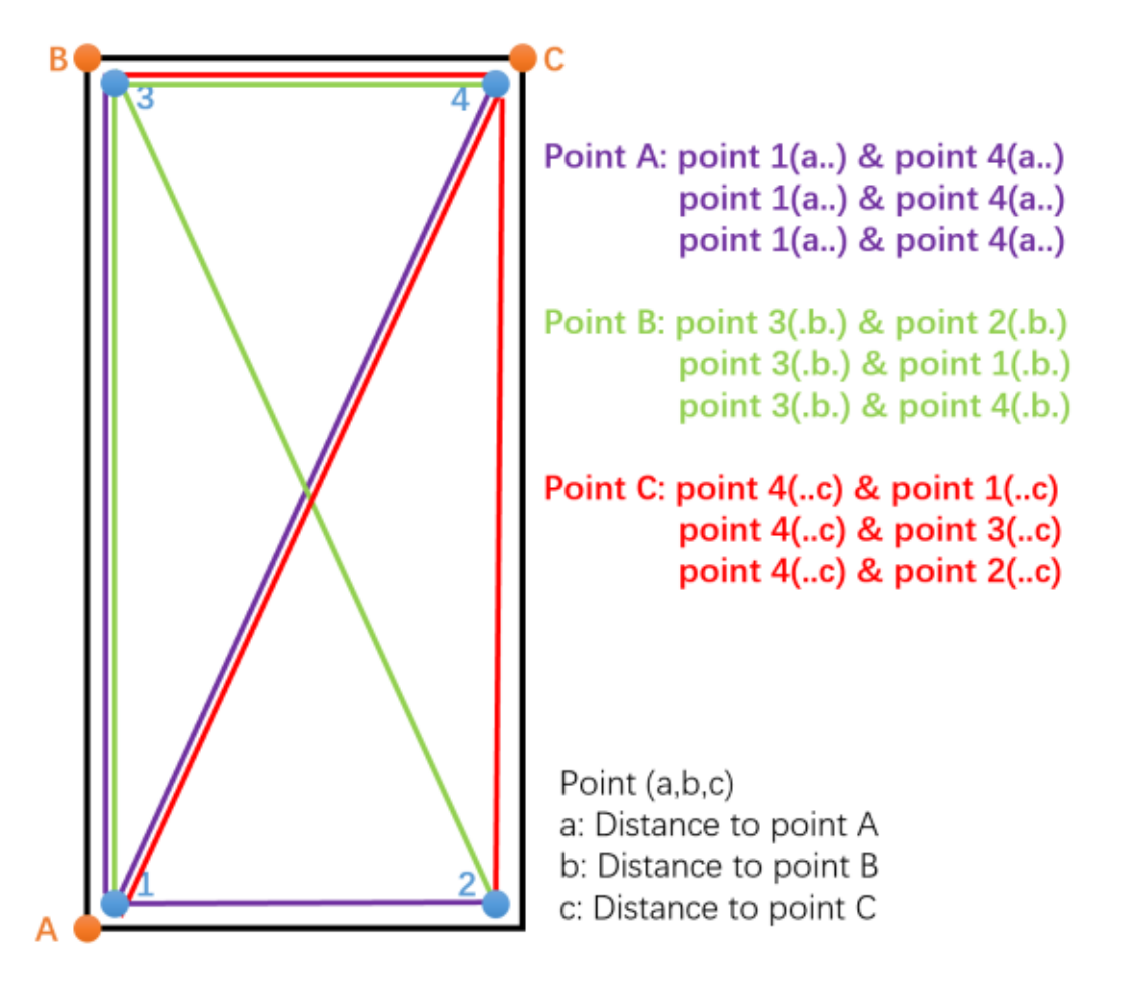}
    \caption{Algorithm diagram of Model Three}
    \label{fig:4.8}
\end{figure}

Based on Model Two, to increase the universality of $a$ and $b$ values, the three-sided corresponding model is established. The specific rules of the three-sided corresponding model are as follows. When calculating $a$ and $b$ values for anchor node A, three sets of distances are used. They are the distances between node A to points 1 and 2, distances to points 1 and 3, and distances to points 1 and 4. Three sets of $a$ and $b$ values are calculated, and the average values of these three $a$ and $b$ values will be used to determine the equation. The same progress is used to calculate the $a$ and $b$ values for anchor node B and anchor node C, as shown in Fig. \ref{fig:4.8}. The data used for node B will be the distances from point 3 to  1, to 2, and to 4. For the anchor node C, it will be calculated by the distances from point  4 to 1, to 2, and to 3. This method will use the average values of $a$ and  $b$ so that it can better reflect the relationship between the actual distance and the tested distance. 

\subsection{Model Four}

\begin{figure}[h!]
    \centering
    \includegraphics[width=8cm]{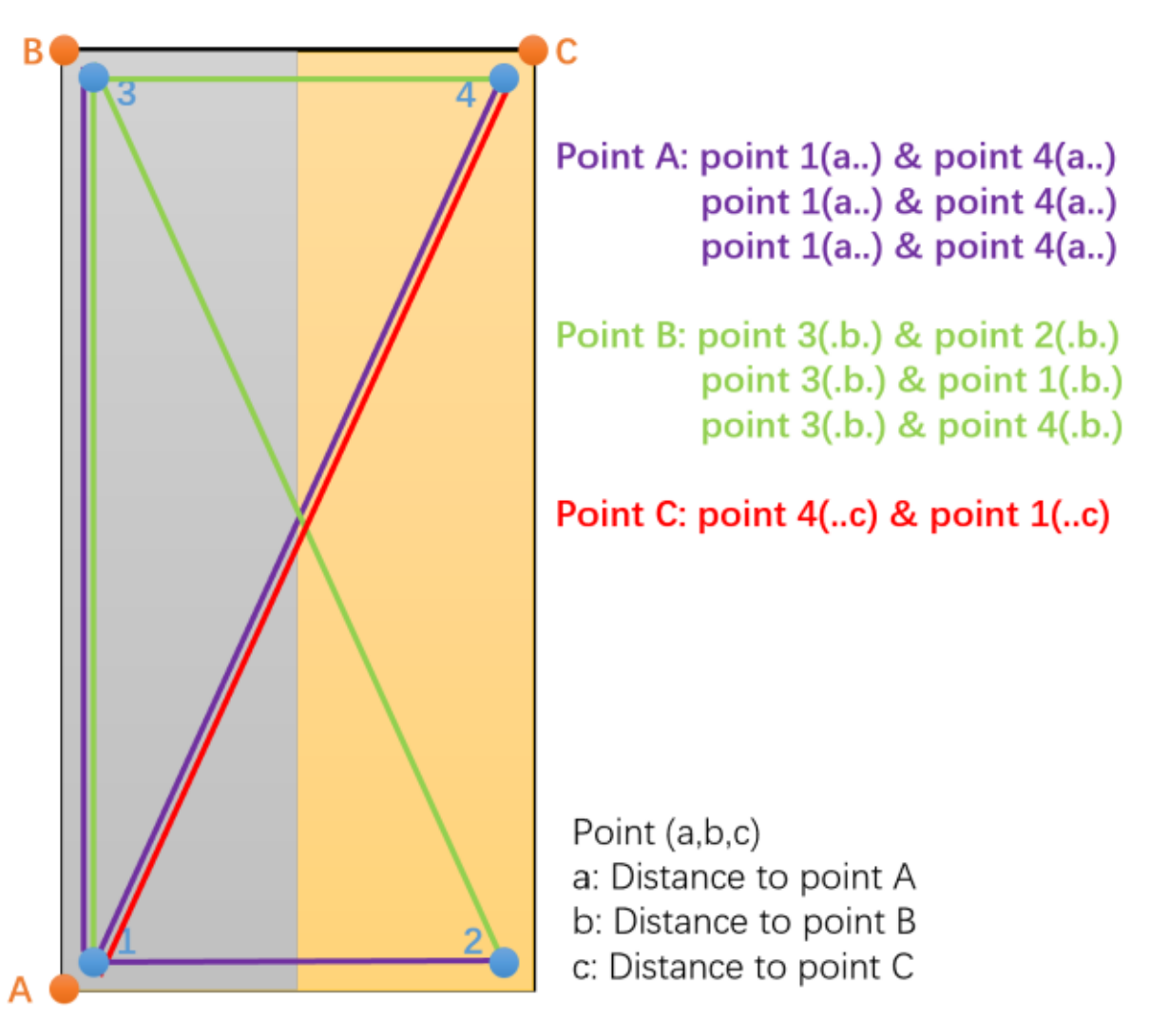}
    \caption{Illustration of Model Four}
    \label{fig:4.9}
\end{figure}

As shown in Fig. \ref{fig:4.9}, there are three anchor nodes, two of them are placed on one side and only one on the other side. On this basis, the observation points on the left will be closer to anchor nodes A and B. Through the analysis of data correction, when the actual distance is less than 1000 mm, the measurement error will be relatively large. Therefore, the use of more averaged $a$ and $b$ values can be helpful to avoid the accidental error. 
In this case, we separate the entire test range into left and right regions. In the left region, we use the method in Model Three to calculate the values of $a$ and $b$. For the anchor node C  in the right region, the $a$ and $b$ values are calculated based on the method of Model Two.

\section{Numerical Results} \label{sec:VII}

\begin{figure}[h!]
    \centering
    \includegraphics[width=6cm]{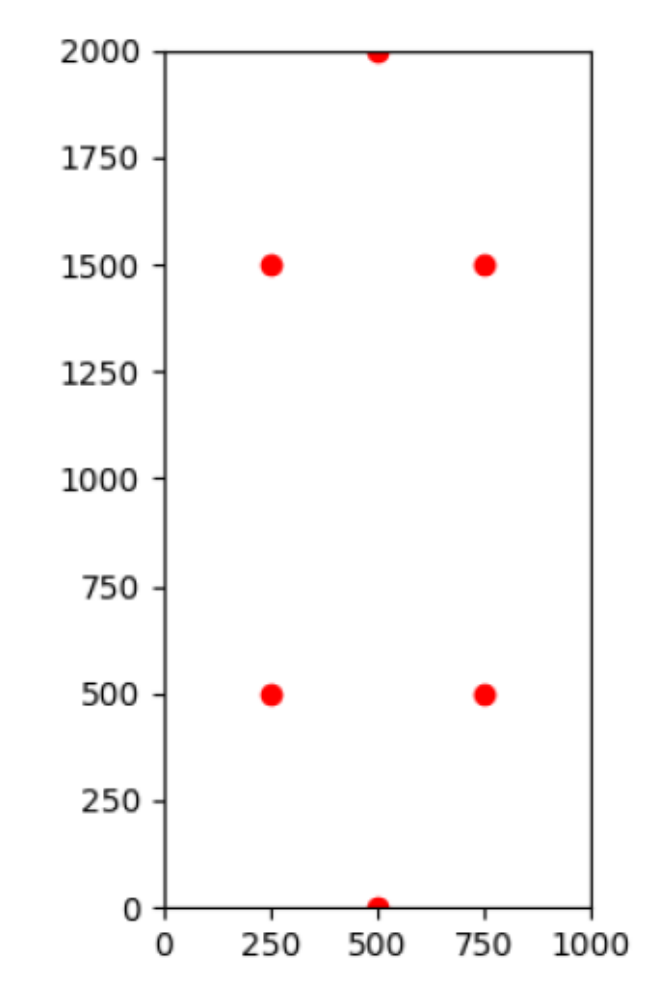}
    \caption{The test point in the experiment.}
    \label{fig:5.1}
\end{figure}

There are six test locations in the experiment, $(250,1500)$, $(250,500)$, $(500,0)$, $(500,2000)$, $(750,1500)$ and $(750,500)$ as shown in Fig. \ref{fig:5.1}. Each of them are tested $400$ times to get an average result.  Among those locations, $(250,1500)$, $(250,500)$, $(750,1500)$ and $(750,500)$ are in the middle of the area and $(500,0)$ and $(500,2000)$ are at the boundary of the area. 

\subsection{Data Processing Result}

\begin{table}[h!]
\renewcommand{\arraystretch}{1.5}
\centering 
\caption{Without machine learning, the average error under different processing methods}
\begin{tabular}{|c c c c c|} 
 \hline
 \multicolumn{5}{|c|}{WITHOUT MACHINE LEARNING (MM)} \\
 \hline
 POINT & 100\% & 90\% &	85\% & 80\% \\  	
 (250,1500) & 358.615 & 178.170 & 115.069 & 99.267 \\ 
 (250,500) & 663.814 & 488.177 & 417.982 & 363.610 \\
 (500,0) & 368.771 & 119.218 & 82.171 & 172.527 \\
 (500,2000) & 553.069 & 257.413 & 128.202 & 67.064 \\
 (750,1500) & 384.831 & 199.945 & 118.857 & 59.392 \\
 (750,500) & 709.586 & 480.109 & 384.500 & 308.022 \\
 \hline
\end{tabular}

\label{table:3}
\end{table}

\begin{table}[h!]
\renewcommand{\arraystretch}{1.5}
\centering
\caption{With machine learning, the average error under different processing methods}
\begin{tabular}{|c c c c c|} 
 \hline
 \multicolumn{5}{|c|}{WITH MACHINE LEARNING (MM))} \\
 \hline
 POINT & 100\% & 90\% &	85\% & 80\% \\  	
 (250,1500) & 213.486 & 211.073 & 215.366 & 213.239 \\ 
 (250,500) & 153.294 & 183.148 & 182.571 & 262.458  \\
 (500,0) & 40.165 & 72.446 & 88.376 & 135.391 \\
 (500,2000) & 35.355 & 31.793 & 90.379 & 105.175  \\
 (750,1500) & 182.718 & 248.962 & 261.501 & 299.888  \\
 (750,500) & 159.202 & 129.999 & 133.952 & 180.624  \\
 \hline
\end{tabular}

\label{table:4}
\end{table}

Tables \ref{table:3} and  \ref{table:4} show the results when using three ratios to correct the raw data. It can be seen from the comparison that without machine learning, the data can be reduced to 80\% of the original to obtain a smaller error. But the data changes too much and the model is biased, so it performs poorly when using machine learning. Therefore, after weighing the two factors, when the measured value is greater than 1000mm, reducing the data to 90\% of the original can achieve relatively better performance for both cases of using or not using  machine learning.

However, this method of data correction also has some problems. The first problem is that it limits the scope of use. In this experiment, the basis for data correction is obtained from observation and analysis of test data. The reliability of this basis is affected by many specific factors, such as the experimental environment, the equipment used in the experiment, the limitation of the size of the experimental scene. If the above factors change, the current data correction methods will not only fail to achieve better performance but may cause larger errors. The second problem is in the data correction process itself. 
It can be seen from the experimental results that when the real distance between the tested location and one of the reference locations is above 1000mm, the measured distance has a large error. 
In this case, 
it is hard to fit the model resulting in poor performance.

\subsection{The Result Without Machine Learning}
\begin{table}[h!]
\renewcommand{\arraystretch}{1.5}
\centering 
\caption{The Result Without Machine Learning}
\begin{tabular}{|c c c|} 
 \hline
  Point & Average error & Maximum error  \\  
  \hline
 (250,1500) & 358.61483 & 628.24358 \\ 
 (250,500) & 663.81404 & 919.35466  \\
 (500,0) & 368.77115 & 438.09246 \\
 (500,2000) & 553.06856 & 627.77862   \\
 (750,1500) & 384.83136 & 522.42224 \\
 (750,500) & 709.58566 & 885.34118  \\
 \hline
\end{tabular}

\label{table:5}
\end{table}

Table \ref{table:5}  shows the maximum error and average error of the six test locations without machine learning. The error is defined as the distance between the calculated location and the real location. The unit used in the experiment is millimeters. 
In the following, we focus on the average error as it is more likely to show the general situation.

\subsection{Results of using Model One}

Table \ref{table:6}  shows the different results in the third module with different machine learning algorithms. There are four algorithms used in this case, namely Decision Tree, Random Forests, KNN and soft voting. The soft voting algorithm determines the location based on results of using the KNN and decision tree with a ratio of $1:2$. 

\begin{table}[h!]
\renewcommand{\arraystretch}{1.5}
\centering
\caption{Average error in Model One with different algorithms}
\begin{tabular}{|c c c c c|} 
 \hline
  Point & Decision Tree	& Random Forests & KNN & Soft Voting\\  
  \hline
 (250,1500)	& 215.28360 & 186.40942 & 332.84053 & 215.28360 \\
 (250,500) & 138.88222 & 212.94731 & 173.30748 & 181.80928 \\
 (500,0) & 34.93499 & 123.06273 & 32.67327 & 33.29100 \\
 (500,2000)	& 28.69175 & 292.12773 & 295.57015 & 28.69175 \\
 (750,1500)	& 193.62215 & 284.71712 & 366.10790 & 195.04524 \\
 (750,500) & 138.73514 & 116.02066 & 64.59321 & 122.41815 \\
 \hline
\end{tabular}
\label{table:6}
\end{table}

As shown in Table \ref{table:6}, the Random Forests algorithm has poor performance compared with the Decision tree and KNN. Therefore, Random Forests is not used for the soft voting. Decision tree has a significantly small error for locations of (250,1500), (500,2000) and (750,1500) and KNN has a clear smaller error only for location of (750,500). In this case, the ratio of the soft voting is set to be $1:2$ for KNN and Decision tree. For all the six test locations, (500,0) and (500,2000) have the smallest error, the distance between the computed test location and real test location is about tens of millimeters.

Compared with the original result without machine learning, the average error is much smaller, indicating that using the investigated machine learning algorithms can increase the accuracy.

\subsection{Three Different Models}

\begin{table}[h!]
\renewcommand{\arraystretch}{1.5}
\centering
\caption{Average error in different Models}
\begin{tabular}{|c c c c|} 
 \hline
  Point & Model One & Model Two & Model Three\\  
  \hline
 (250,1500)	& 215.28360 & 186.36835 & 110.22056  \\
 (250,500) & 181.80928 & 238.18274 & 174.05819 \\
 (500,0) & 33.29100 & 41.50481 & 180.06017 \\
 (500,2000)	& 28.69175 & 119.21513 & 311.41235 \\
 (750,1500)	& 195.04524 & 121.03428 & 183.53073  \\
 (750,500) & 122.41815 & 52.89492 & 50.92223\\
 \hline
\end{tabular}

\label{table:7}
\end{table}

\begin{figure}[h!]
    \centering
    \includegraphics[width=9cm]{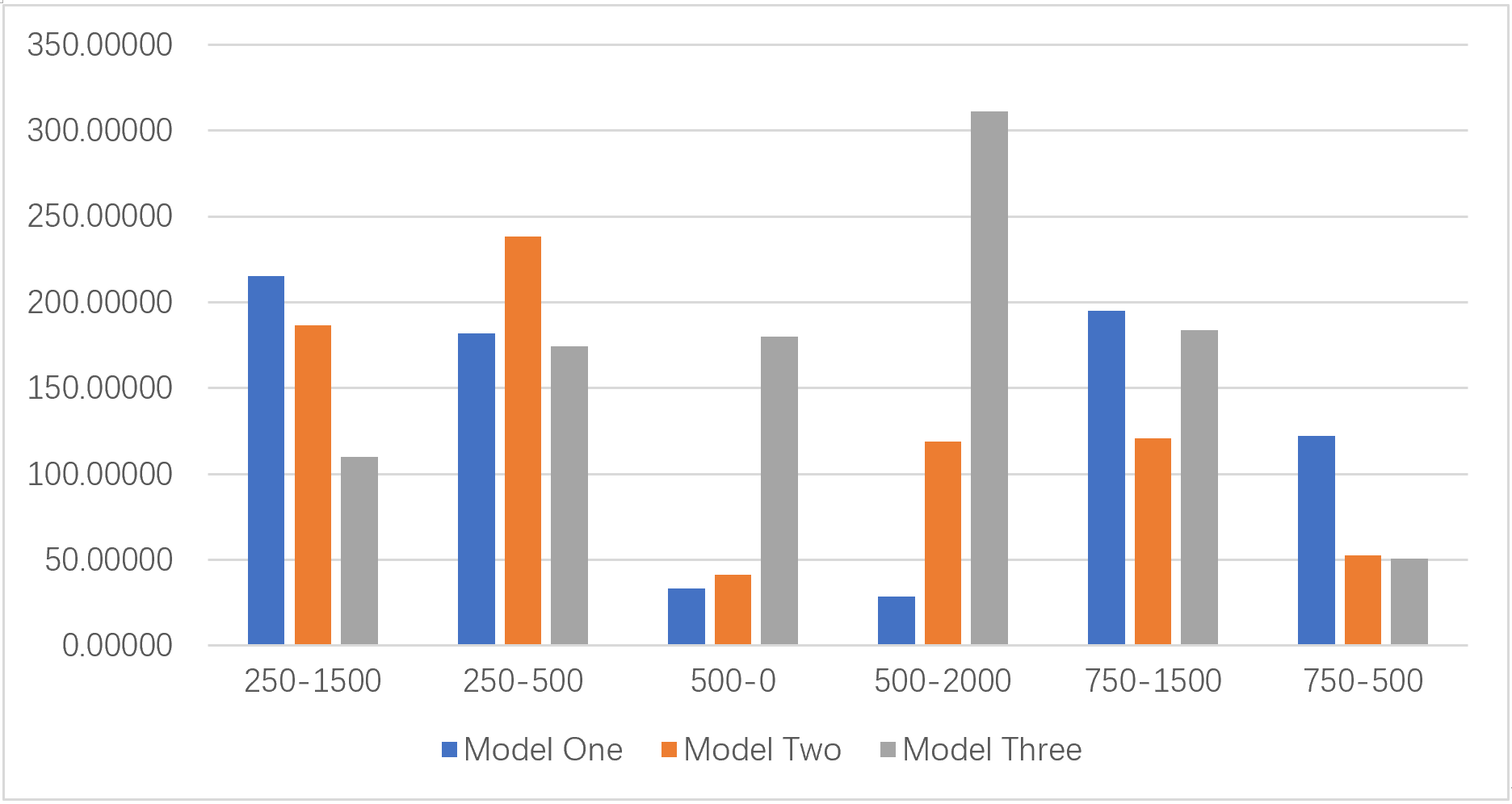}
    \caption{Average error in different Models.}
    \label{fig:5.2}
\end{figure}

Table \ref{table:7}  and Fig. \ref{fig:5.2} show the results when using the Soft Voting algorithm and training in three different models.  Worth to mention that for locations $(250,1500)$ and $(250,500)$, Model three has the smallest average error, for locations $(500,0)$ and $(500,2000)$, Model one gives the smallest average error. And at locations $(750,500)$ and $(750,1500)$, the average error using the second model is much smaller than others. In this case, it is possible to get better performance by combining different models.

\subsection{Result in Model Four}

Table  \ref{table:8} shows results of the third model using different machine learning algorithms. The algorithms used in this case are KNN and soft voting. In the experiment, unlike the previous model, KNN algorithm has a better performance than the Decision tree algorithm. So, the soft voting algorithm consists of the KNN and decision tree at a ratio of $3:1$. 

\begin{table}[h!]
\renewcommand{\arraystretch}{1.5}
\centering
\caption{Average error in Model Four with KNN and soft voting.}
\begin{tabular}{|c c c|} 
 \hline
  Point & KNN & Soft Voting\\
  \hline
(250,1500) & 102.57647 & 97.82208 \\
(250,500) & 259.96927 & 288.14183 \\
(500,0) & 168.91243 & 142.42520 \\
(500,2000) & 25.00000 & 25.00000 \\
(750,1500) & 114.51841 & 119.51332 \\
(750,500) & 53.50840 & 53.57558\\
 \hline
\end{tabular}

\label{table:8}
\end{table}

\begin{figure}[h]
    \centering
    \includegraphics[width=9cm]{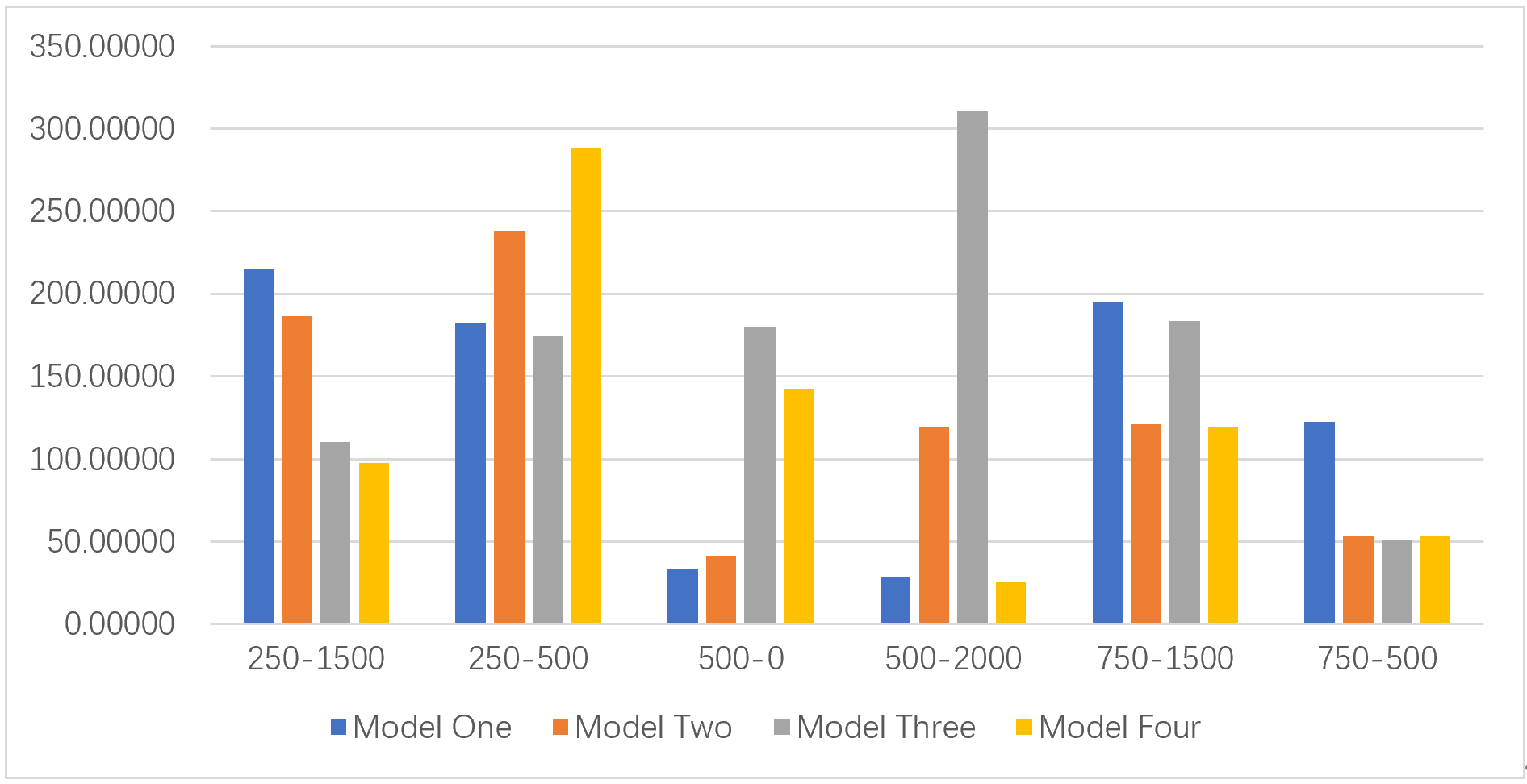}
    \caption{Average error in different Models with soft voting.}
    \label{fig:5.3}
\end{figure}

Fig. \ref{fig:5.3} shows the average error performance for all the four models, it is clear that the fourth model could achieve high accuracy for most of test locations, except for location (250,500).

\subsection{Comparison of Experimental Results With or Without Machine Learning}

\begin{figure}[h!]
    \centering
    \includegraphics[width=9cm]{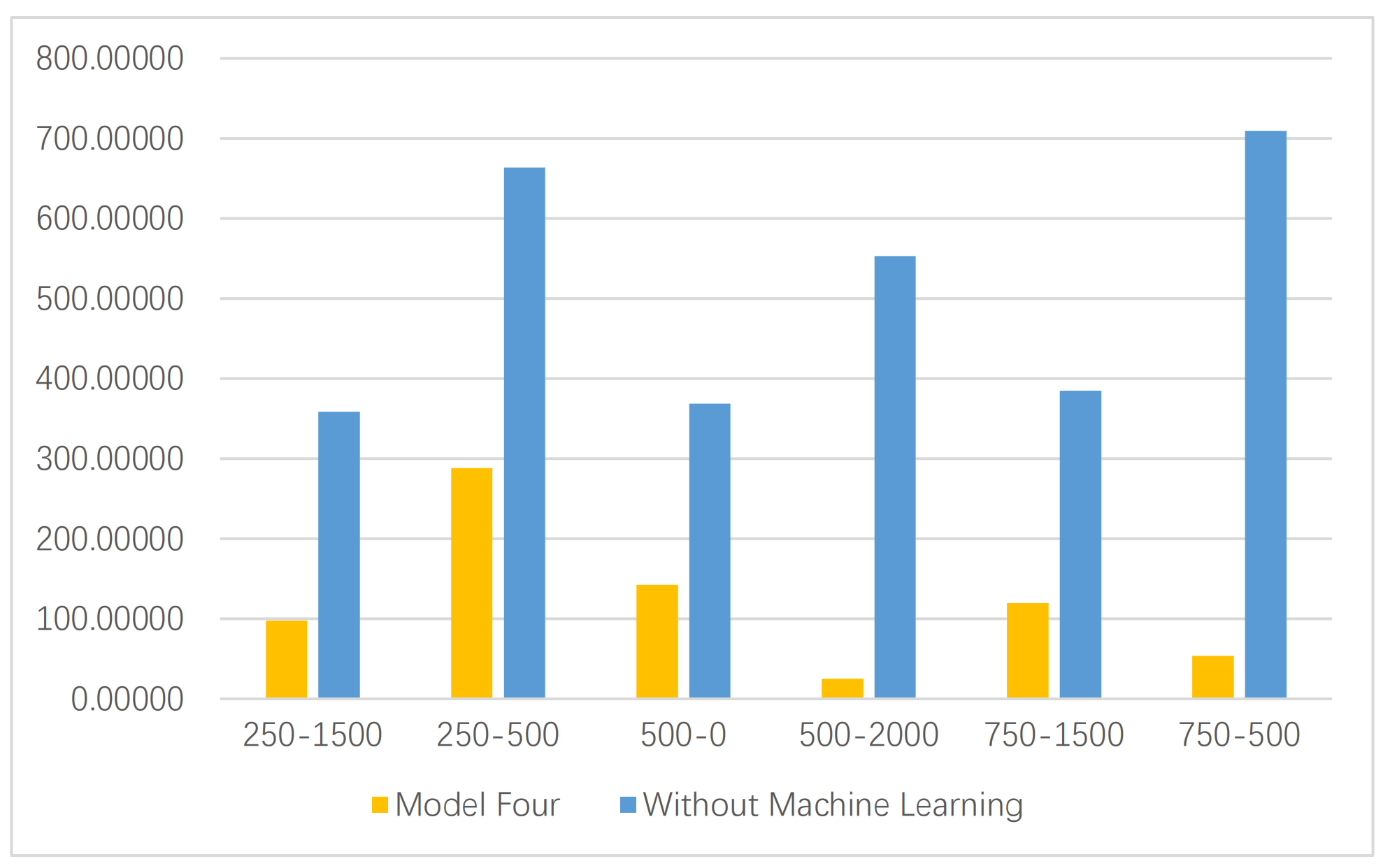}
    \caption{Average error in Model Four and Average error without machine learning}
    \label{fig:5.4}
\end{figure}

As shown in Fig.\ref{fig:5.4}, when using machine learning algorithms, the average error is significantly reduced. The performance is most obvious at the test location of $(500, 2000)$, with an average error reducing from $553$ mm to $25$ mm. At the location of $(250, 500)$, the reduction is small, only reduced from 663 mm to 288 mm. This poor performance may be due to the large error in the test process, which makes it impossible to match the model well.

\section{Conclusion}\label{sec:VIII}

This study describes a Machine Learning based indoor positioning system. A formula is used to compute the data while creating a model for machine learning using fingerprint positioning. A considerable number of measurements can be avoided this way. The report compares and contrasts the performance of four distinct models. Most test location' average inaccuracy can be reduced to less than 150 mm using the best model. For the future work, we plan to use deep learning algorithms and denoising techniques developed in our existing works, e.g., \cite{IREALCARE1,IREALCARE2,IREALCARE3,IREALCARE4,IREALCARE5} to improve the accuracy. Privacy concerns for location-based services \cite{privacy1,privacy2,privacy3,privacy4} and cellular networks \cite{NC,JNCC, UAV_THz} and sensor networks \cite{NC4,WRN} based location techniques are also our future plan. 






\end{document}